\documentclass{appolb}
\usepackage{graphicx}
\usepackage{amsmath}
\usepackage{epsfig}
\usepackage{pstricks,pst-node,pst-text,pst-3d}
\usepackage{psfrag}
\usepackage[square,comma,sort&compress]{natbib}

\newcommand {\ignore}[1]{}

\def\slash#1{#1\!\!\! /}

\def\ifmath#1{\relax\ifmmode #1\else $#1$\fi}
\def\half{\ifmath{{\textstyle{\frac{1}{2}}}}}

\def\ds{\displaystyle}
\def\beqa{\begin{eqnarray}}
\def\eeqa{\end{eqnarray}}
\def\ra{\rightarrow}
\def\Fig#1{Fig.~\ref{#1}}

\def\te{{\tilde e}}
\def\tm{{\tilde \mu}}
\def\st{{\tilde \tau}}

\def\rp{$R_p \hspace{-1em}/\;\:$ }

\def\bold#1{\setbox0=\hbox{$#1$} 
     \kern-.025em\copy0\kern-\wd0 
     \kern.05em\copy0\kern-\wd0 
     \kern-.025em\raise.0433em\box0 }
\newcommand{\bemartin}[1]{\begin{equation} \label{(#1)}}
\newcommand{\eemartin}{\end{equation}}
\newcommand{\bamartin}[1]{\begin{eqnarray} \label{(#1)}}
\newcommand{\eamartin}{\end{eqnarray}}

\DeclareMathAlphabet{\mathsc}{OT1}{cmr}{m}{sc}
\newcommand{\Sol}  {\mathsc{sol}}

\newcommand{\Atm}  {\mathsc{atm}}

\newrgbcolor{orange}{0.97 0.75 0.05}
\newrgbcolor{istblue}{0.47 0.74 1}
\newrgbcolor{darkgreen}{0.13 0.50 0.09}

\def\21{SU(2) $\otimes$ U(1) }
\def\vev#1{\left\langle #1\right\rangle}
\newcommand{\FIGURE}[1]{\begin{figure}[ht] {#1} \end{figure}}
\def\Eq#1{Eq. (\ref{#1})}

\begin{document}
\title{Neutrino Properties in Supersymmetric Models with R-parity Breaking
\thanks{Based on talks given at the Symposium in Honour of Gustavo
  C. Branco and at the Corfu Summer Institute on EPP.}%
}
\author{Jorge C. Rom\~ao
\address{Instituto Superior T\'ecnico, Departamento de F\'{\i}sica\\
             A. Rovisco Pais 1, 1049-001 Lisboa, Portugal}
}
\maketitle
\begin{abstract}
We review supersymmetric models where R--parity is broken either
explicitly or spontaneously.  The simplest unified extension of the
\texttt{MSSM} with explicit bilinear R--Parity violation provides a predictive
scheme for neutrino masses and mixings which can account for the
observed atmospheric and solar neutrino anomalies.  Despite the
smallness of neutrino masses R-parity violation is observable at
present and future high-energy colliders, providing an unambiguous
cross-check of the model. This model can be shown to be an effective
model for the, more theoretically satisfying, spontaneous broken
theory. The main difference in this last case is the appearance of a
massless particle, the majoron, that can modify the decay modes of the
Higgs boson, making it decay invisibly most of the time.
\end{abstract}
  
\section{Introduction}

Despite the tremendous effort that has led to the discovery of
neutrino mass~\cite{fukuda:1998mi,ahmad:2002jz,eguchi:2002dm} the
mechanism of neutrino mass generation will remain open for years to
come (a detailed analysis of the three--neutrino oscillation
parameters can be found in ~\cite{maltoni:2004ei}).  The most popular
mechanism to generate neutrino masses is the seesaw
mechanism~\cite{gell-mann:1980vs,yanagida:1979,mohapatra:1981yp,chikashige:1981ui,schechter:1980gr}.
Although the seesaw fits naturally in SO(10) unification models, we
currently have no clear hints that uniquely point towards any
unification scheme.

Therefore it may well be that neutrino masses arise from physics
having nothing to do with unification, such as certain seesaw
variants~\cite{mohapatra:1986bd}, and models with radiative
generation~\cite{zee:1980ai,babu:1988ki}.
Here we focus on the specific case of low-energy supersymmetry with
violation of R--parity, as the origin of neutrino
mass. R--parity is defined as $R_p = (-1)^{3B+L+2S}$ with $S$,
$B$, $L$ denoting spin, baryon and lepton numbers,
respectively~\cite{aulakh:1982yn}.

In these models R--parity can be broken either explicitly or
spontaneously. In the first case we consider the bilinear R--parity
violation model, the simplest effective description of R-parity
violation~\cite{diaz:1998xc}. The model not only accounts for the
observed pattern of neutrino masses and
mixing~\cite{diaz:2003as,hirsch:2000ef,romao:1999up,hirsch:2004he},
but also makes predictions for the decay branching ratios of the
lightest supersymmetric
particle~\cite{hirsch:2002ys,porod:2000hv,restrepo:2001me,hirsch:2003fe}
from the current measurements of neutrino mixing
angles~\cite{maltoni:2004ei}. In the second case R--parity violation
takes place ``a la Higgs'', i.e., spontaneously, due to non-zero
sneutrino vacuum expectation values
(vevs)~\cite{masiero:1990uj,romao:1992vu,shiraishi:1993di}.
In this case one of the neutral CP-odd scalars is identified with the
majoron, $J$. In contrast with the seesaw majoron, ours is characterized by
a small scale (TeV-like) and carries only one unit of lepton number.
In previous
studies~\cite{romao:1992ex,romao:1992zx,decampos:1994rw,decampos:1996bg}
it was noted that the spontaneously broken R--parity (\texttt{SBRp})
model leads to the possibility of invisibly decaying Higgs bosons,
provided there is an \21 singlet superfield $\Phi$ coupling to the
electroweak doublet Higgses, the same that appears in the
\texttt{NMSSM}. We have reanalyzed \cite{hirsch:2004rw} this issue
taking into account the small masses indicated by current neutrino
oscillation data~\cite{maltoni:2004ei}.  We have shown explicitly that
the invisible Higgs boson decay Eq.~(\ref{eq:HJJ}),
\begin{equation}
  \label{eq:HJJ}
  h \to J J
\end{equation}
can be the most important mode of Higgs boson decay. This
is remarkable, given the smallness of neutrino masses required to fit
current neutrino oscillation data.

\section{Bilinear R-Parity Violation}

\subsection{The Model}

Since \texttt{BRpV} has been discussed in the literature several times
\cite{romao:1999up,hirsch:2000ef,diaz:1998xc,decampos:1995av,
akeroyd:1998iq,banks:1995by} we will repeat only the main features of
the model here.  We will follow the notation of
\cite{romao:1999up,hirsch:2000ef}.  The simplest bilinear \rp model
(we call it the \rp \texttt{MSSM}) is characterized by the
superpotential
\begin{equation}
\label{eq:Wpot} 
W = W_{\rm MSSM} + W_{\slash R_P} 
\end{equation} 
In this equation $W_{\rm MSSM}$ is the ordinary superpotential of the
\texttt{MSSM},
\begin{equation}
W=\varepsilon_{ab}\left[
 h_U^{ij}\widehat Q_i^a\widehat U_j\widehat H_u^b
+h_D^{ij}\widehat Q_i^b\widehat D_j\widehat H_d^a
+h_E^{ij}\widehat L_i^b\widehat R_j\widehat H_d^a 
-\mu\widehat H_d^a\widehat H_u^b \right]
\end{equation}
where $i,j=1,2,3$ are generation indices, $a,b=1,2$ are $SU(2)$
indices. We have three additional terms that break R-parity,
\begin{equation}
\label{eq:WRPV} 
W_{\slash R_P} = \epsilon_{ab} \epsilon_i \widehat
  L^a_i\widehat H^b_u.  
\end{equation} 
These bilinear terms, together with the corresponding terms in the
soft supersymmetric (\texttt{SUSY}) breaking part of the Lagrangian, 
\begin{equation}
\label{eq:Lsoft}
  V_{\rm soft} = V_{\rm soft}^{\rm MSSM} +
  \epsilon_{ab} B_i \epsilon_i {\tilde L}^a_i H^b_u 
\end{equation} 
define the minimal model, which we will adopt throughout this paper.
The appearance of the lepton number violating terms in
Eq.~(\ref{eq:WRPV}) leads, in general, to non-zero vacuum expectation
values for the scalar neutrinos $\langle {\tilde \nu}_i \rangle$,
called $v_i$ in the rest of this paper, in addition to the VEVs $v_u$
and $v_d$ of the \texttt{MSSM} Higgs fields $H_u^0$ and $H_d^0$.
Together with the bilinear parameters $\epsilon_i$, the $v_i$ induce
mixing between various particles which in the \texttt{MSSM} are
distinguished (only) by lepton number (or R--parity).  Mixing between
the neutrinos and the neutralinos of the \texttt{MSSM} generates a
non-zero mass for one specific linear superposition of the three
neutrino flavor states of the model at tree-level while 1-loop
corrections provide mass for the remaining two neutrino
states~\cite{romao:1999up,hirsch:2000ef}.

\subsection{Tree Level Neutral Fermion Mass Matrix}

In the basis $\psi^{0T}=
(-i\lambda',-i\lambda^3,\widetilde{H}_d^1,\widetilde{H}_u^2, \nu_{e},
\nu_{\mu}, \nu_{\tau} )$ the neutral fer\-mions mass terms in the
Lagrangian are given by
\begin{equation}
\mathcal{L}_m=-\frac 12(\psi^0)^T{\bold M}_N\psi^0+h.c.   
\end{equation}
where the neutralino/neutrino mass matrix is 
\begin{equation}
{\bold M}_N=\left[  
\begin{array}{cc}  
\mathcal{M}_{\chi^0}& m^T \cr
m & 0 \cr
\end{array}
\right]
\end{equation}
with
\begin{equation}
\mathcal{M}_{\chi^0}=\left[  
\begin{array}{cccc}  
M_1 & 0 & -\frac 12g^{\prime }v_d & \frac 12g^{\prime }v_u \cr
0 & M_2 & \frac 12gv_d & -\frac 12gv_u \cr
-\frac 12g^{\prime }v_d & \frac 12gv_d & 0 & -\mu  \cr
\frac 12g^{\prime }v_u & -\frac 12gv_u &  -\mu & 0  \cr
\end{array}  
\right] 
\quad ; \quad
m=\left[  
\begin{array}{c}
a_1 \cr
a_2 \cr
a_3 
\end{array}  
\right] 
\end{equation}
where $a_i=(-\frac 12g^{\prime }v_i, \frac 12gv_i, 0,\epsilon_i)$. 
This neutralino/neutrino mass matrix is diagonalized by 
\begin{equation}
\mathcal{ N}^*{\bold M}_N\mathcal{N}^{-1}={\rm diag}(m_{\chi^0_1},m_{\chi^0_2}, 
m_{\chi^0_3},m_{\chi^0_4},m_{\nu_1},m_{\nu_2},m_{\nu_3}) 
\label{eq:NeuMdiag} 
\end{equation}

\subsubsection{Approximate Diagonalization at Tree Level }

If the \rp parameters are small (we will show below that this is
indeed the case), then we can block-diagonalize ${\bold M}_N$
approximately to the form diag($m_{\rm eff},\mathcal{M}_{\chi^0}$)
\begin{equation}
m_{\rm eff} = - m \cdot \mathcal{M}_{\chi^0}^{-1} m^T = 
\frac{M_1 g^2 + M_2 {g'}^2}{4\, \det(\mathcal{M}_{\chi^0})} 
\left(\begin{array}{ccc}
\Lambda_e^2 & \Lambda_e \Lambda_\mu
& \Lambda_e \Lambda_\tau \\
\Lambda_e \Lambda_\mu & \Lambda_\mu^2
& \Lambda_\mu \Lambda_\tau \\
\Lambda_e \Lambda_\tau & \Lambda_\mu \Lambda_\tau & \Lambda_\tau^2
\end{array}\right),
\end{equation}
with
\begin{equation}
\label{eq:lambda}
\Lambda_i=\mu v_i + v_d \epsilon_i \, .
\end{equation}
The matrices $N$ and $V_{\nu}$ diagonalize 
$\mathcal{M}_{\chi^0}$ and $m_{\rm eff}$ 
\begin{equation}
N^{*}\mathcal{M}_{\chi^0} N^{\dagger} = {\rm diag}(m_{\chi^0_i})
\quad ; \quad
V_{\nu}^T m_{\rm eff} V_{\nu} = {\rm diag}(0,0,m_{\nu}),
\end{equation}
where 
\begin{equation}
m_{\nu} = Tr(m_{\rm eff}) = 
\frac{M_1 g^2 + M_2 {g'}^2}{4\, \det(\mathcal{M}_{\chi^0})} 
|{\vec \Lambda}|^2.
\label{eq:mneu3}
\end{equation}

So we get at tree level only one massive neutrino, the two other
eigenstates remaining massless. The tree level value will give the
atmospheric mass scale, while the other states will get mass at one
loop level. Therefore the \texttt{BRpV} model produces a hierarchical mass 
spectrum. We will show below how the one loop masses are generated.

\subsection{One--Loop Neutrino Masses and Mixings}

\subsubsection{Definition}

The self--energy for the neutralino/neutrino is~\cite{hirsch:2000ef},

\begin{equation}
\hskip 3.5cm \equiv
i \left\{ \slash{p} \left[ P_L \Sigma^L_{ij} + P_R \Sigma^R_{ij} \right]
-\left[ P_L \Pi^L_{ij} + P_R \Pi^R_{ij} \right]\right\}
\end{equation}

\begin{picture}(0,0)
\put(-0.5,10){\includegraphics[width=2.7cm]{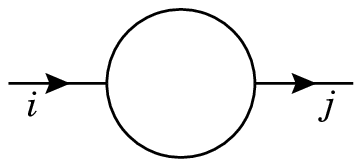}}
\end{picture}

\noindent
Then the pole mass is,
\begin{equation}
M^{\rm pole}_{ij}= M^{\rm \overline{DR}}_{ij}(\mu_R) + \Delta M_{ij}
\end{equation}
with
\begin{equation}
\Delta M_{ij}\! =\! \left[ \half 
\left(\Pi^V_{ij}(m_i^2)\! +\! \Pi^V_{ij}(m_j^2)\right) 
\!-\! \half 
\left( m_{\chi^0_i} \Sigma^V_{ij}(m_i^2) \! +\! 
m_{\chi^0_j} \Sigma^V_{ij}(m_j^2) \right) \right]_{\Delta=0}
\end{equation}
where
\begin{equation}
\Sigma^V=\half \left(\Sigma^L+\Sigma^R\right)
\quad ; \quad
\Pi^V=\half \left(\Pi^L+\Pi^R\right)
\end{equation}
an the parameter that appears in dimensional reduction is,
\begin{equation}
\ds \Delta=\frac{2}{4-d} -\gamma_E + \ln 4\pi \ .
\end{equation}
Explicit expressions can be found in~\cite{hirsch:2000ef}.

\subsubsection{One--loop results for the masses}

\begin{figure}[htbp]
  \centering
\begin{tabular}{cc}
  \psfrag{x}{$\tiny \epsilon^2/|\vec \Lambda|$}
  \psfrag{y}{$\tiny m_{ 1,2,3}$ {\small (eV)}}
  \includegraphics[clip,height=55mm]{masses-corfu2005.eps}
  &
  \includegraphics[clip,height=55mm]{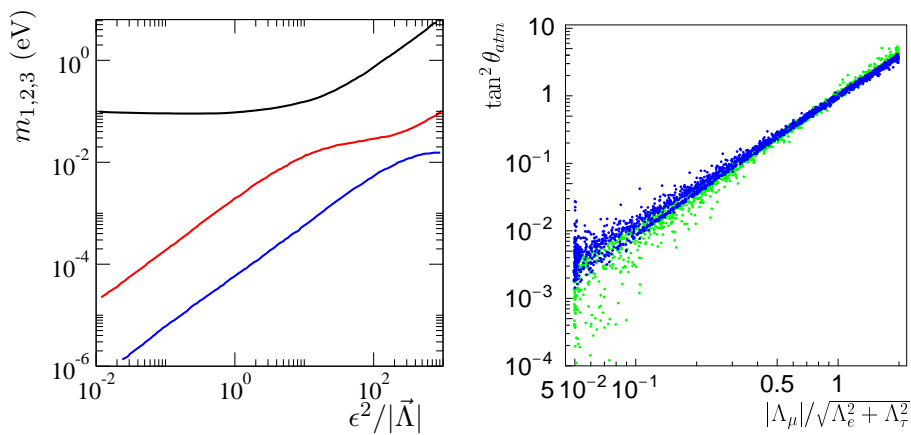}
\end{tabular}
\caption{\small 
 a) One-loop neutrino masses.
  b) Atmospheric angle as a function of $|\Lambda_{\mu}| /
  \sqrt{\Lambda^2_{e}+\Lambda^2_{\tau}}$. } 
\label{fig:numasses}
\end{figure}

The \texttt{BRpV} model produces a hierarchical mass spectrum for
almost all choices of parameters. The largest mass can be estimated by
the tree level value using Eq.~(\ref{eq:mneu3}).  Correct $\Delta
m^2_{\Atm}$ can be easily obtained by an appropriate choice of $| \vec
\Lambda|$. The mass scale for the solar neutrinos is generated at
1--loop level and therefore depends in a complicated way in the model
parameters. We will see below how to get an approximate formula for
the solar mass valid for most cases of interest.  Here we just present
in Fig.~\ref{fig:numasses} a), for illustration purposes, the plot of
the three eigenstates as a function of the parameter
$\epsilon^2/|\Lambda|$, for particular values of the \texttt{SUSY}
parameters, $m_0=\mu=500$ GeV, $\tan \beta= 5$, $B=-A=m_0$. For the
R-parity parameters we took $|\vec \Lambda|=0.16$ GeV, $10*
\Lambda_e=\Lambda_{\mu}=\Lambda_{\tau}$ and
$\epsilon_1=\epsilon_2=\epsilon_3$.

\subsubsection{The mixings}

\begin{figure}[ht]
\centering
\begin{tabular}{cc}
\includegraphics[width=0.45\textwidth]{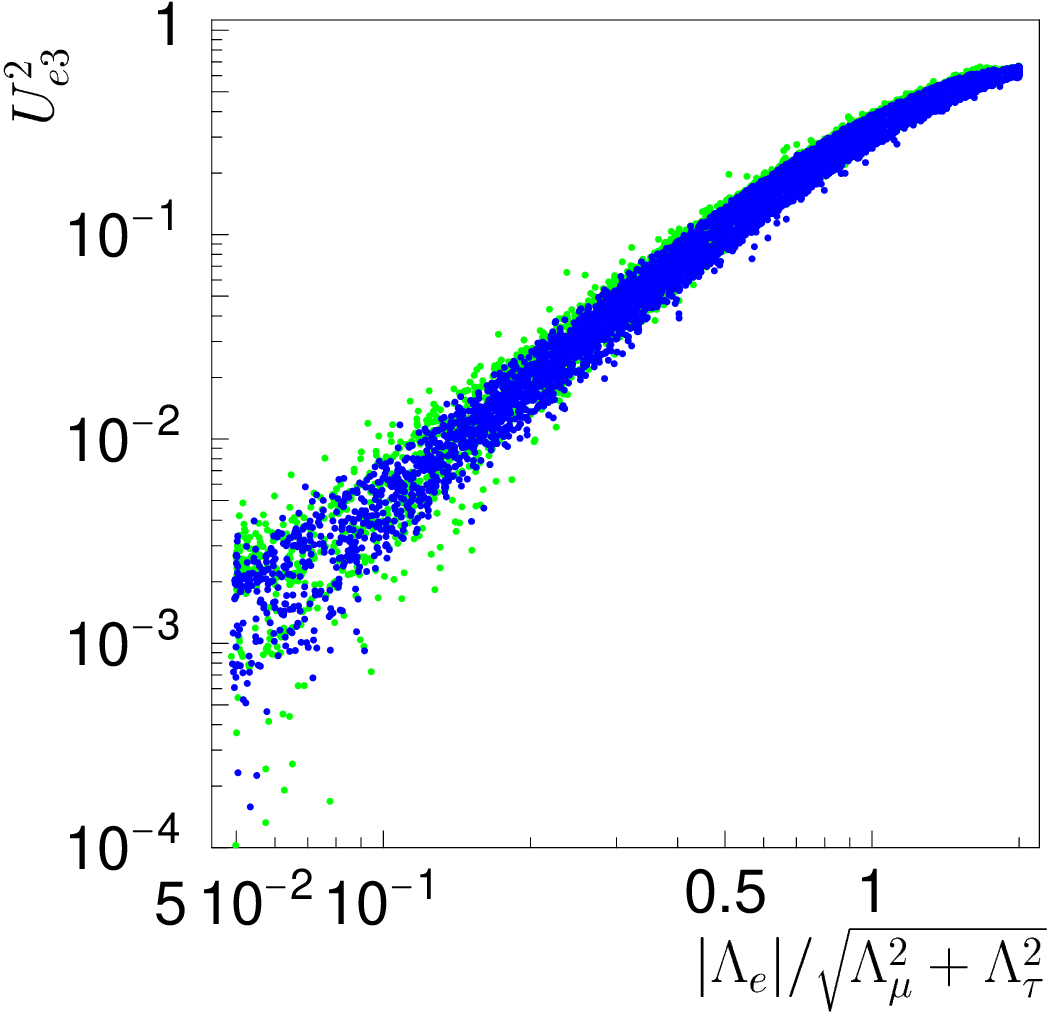}
&
\includegraphics[width=0.45\textwidth]{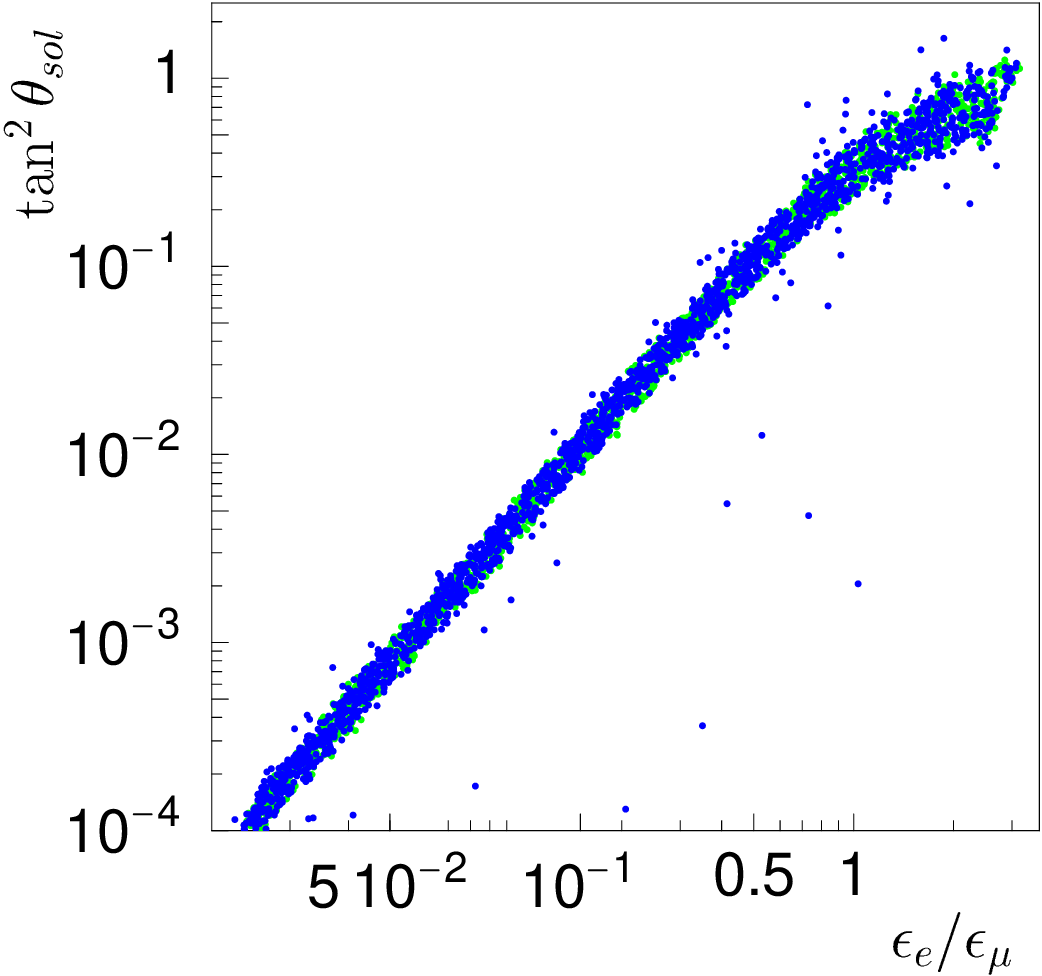}
\end{tabular}
\vspace{-2mm}
\caption{\small 
a) Atmospheric angle as a function of
  $|\Lambda_{\mu}|/\sqrt{\Lambda^2_{e}+\Lambda^2_{\tau}}$. 
  b) $U^2_{e3}$ as a function of
  $|\Lambda_e|/\sqrt{\Lambda^2_{\mu}+\Lambda^2_{\tau}}$.  
}
\label{corfu_fig2}
\end{figure}

Now we turn to the discussion of the mixing angles.  As can be seen
from \Fig{fig:numasses} a), if $\epsilon^2/|\vec \Lambda| \ll 100$, then
the 1--loop corrections are not larger than the tree level results. In
this case the flavor composition of the 3rd mass eigenstate is
approximately given by
\begin{equation}
U_{\alpha 3}\approx\Lambda_{\alpha}/|\vec \Lambda |
\end{equation}
As the atmospheric and reactor neutrino data tell us that
$\nu_{\mu}\ra \nu_{\tau}$ oscillations are preferred over 
$\nu_{\mu}\ra \nu_e$, we conclude that  
\begin{equation}
\Lambda_e \ll \Lambda_{\mu} \simeq \Lambda_{\tau}
\end{equation}
are required for \texttt{BRpV} to fit the data.  We cannot set all the
$\Lambda_i$ equal, because in this case $U_{e 3}$ would be too large
contradicting the \texttt{CHOOZ}, as shown in Fig.~\ref{corfu_fig2}
a).

We have shown that even a very small deviation from
universality (less then 1\%) of the soft parameters at the
\texttt{GUT} scale allows for
\begin{equation}
\frac{\epsilon_e}{\epsilon_{\mu}}\not=\frac{\Lambda_e}{\Lambda_{\mu}}  
\end{equation}
Then we can have at the same time \textbf{small} $U_{e3}^2$ determined
by $\Lambda_e/\Lambda_{\mu}$ as in Fig.~\ref{corfu_fig2} a) and
\textbf{large} $\tan^2(\theta_{\Sol})$ determined by
$\epsilon_e/\epsilon_{\mu}$ as in Fig.~\ref{corfu_fig2} b). After the
\texttt{Kamland} and \texttt{SNO} salt results, this is the only scenario
consistent with the data.

\subsection{Approximate formulas for the solar mass and mixing}

In all the previous analysis we used a numerical program to evaluate
the one-loop masses and mixings. It is however desirable to have
analytical approximate results that can give us quickly the most
important contributions.  We have identified that these are the 
bottom-sbottom loops and the charged scalar-charged fermion loops.
Then, by expanding in powers of the small R-Parity breaking parameters
$\epsilon_i$, we get approximate formulas for the solar mass scale and 
mixing angle as we will explain below.

\subsubsection{Bottom-sbottom loops}

As an example, we give the result from the bottom-sbottom
loop \cite{diaz:2003as},  
\begin{eqnarray}
\Delta M_{ij}&=&-{\frac{N_c m_b}{16\pi^2}}
2s_{\tilde b}c_{\tilde b}h_b^2
\Delta B_0^{\tilde b_1\tilde b_2}
\left[
\frac{\tilde\epsilon_i\tilde\epsilon_j}{\mu^2}
  + a_3 b \left(\tilde\epsilon_i\delta_{j3}+
 \tilde\epsilon_j\delta_{i3}\right)|\vec\Lambda| \right. \nonumber \\
 &&\hskip 40mm \left. + \left( a_3^2 + \frac{ a_L a_R}{h^2_b}\right)
 \delta_{i3}\delta_{j3} 
 |\vec\Lambda|^2 \right] \, ,
\label{eq:bot-sbot}
 \end{eqnarray}
where $\widetilde \epsilon_i$ are the $\epsilon_i$ in the basis where
the tree level neutrino mass matrix is diagonal,
\begin{equation}
\widetilde \epsilon_i = 
\left(V_\nu^{(0)T}\right)^{ij} \epsilon_j \, ,
\end{equation}
the $a_i$ are functions of the \texttt{SUSY} parameters, and
\begin{equation}
  \label{eq:3}
 \Delta B_0^{\tilde b_1\tilde b_2}=
  B_0(0,m_b^2,m_{\tilde b_1}^2)-B_0(0,m_b^2,m_{\tilde b_2}^2)\, .
  \end{equation}
where the $B_0$ are the Passarino--Veltman loop
functions~\cite{passarino:1979jh}. 
The different contributions can be understood as coming from different
types of insertions as shown in \Fig{fig:botsbot}. In this figure 
open circles correspond to small R-parity violating projections, 
full circles to R-parity conserving projections, and
open circles with a cross inside to mass insertions which flip
chirality. With this understanding one can make a one to one
correspondence between Eq.~(\ref{eq:bot-sbot}) and \Fig{fig:botsbot}.
\FIGURE{
  \begin{tabular}{cc}
    \includegraphics[width=0.45\linewidth]{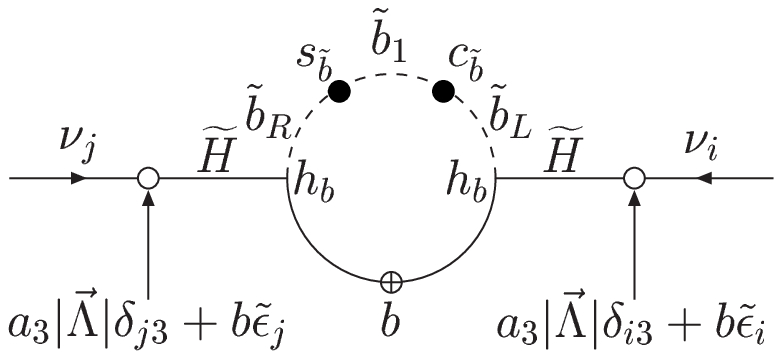}
    &\includegraphics[width=0.45\linewidth]{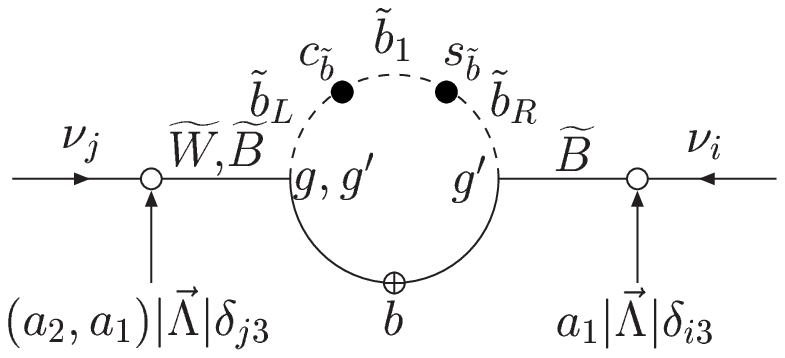}
  \end{tabular}
  \caption{Bottom-sbottom loop and different types of insertions.}
  \label{fig:botsbot}
}

\subsubsection{Simplified approximation formulas}

\begin{figure}[htbp]
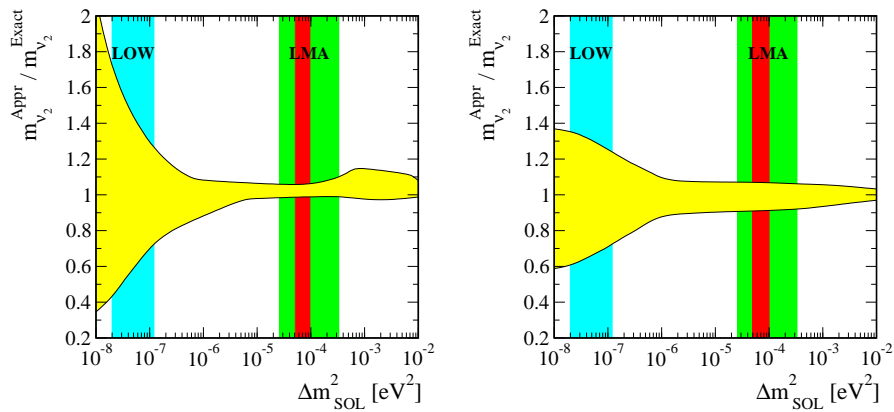

\centering
\begin{tabular}{cc}
  \includegraphics[width=0.45\linewidth]{Nf2002Sim-v5.eps}
  &\includegraphics[width=0.45\linewidth]{StauSim-v5.eps}
\end{tabular}
\caption{Comparison between the simplified analytical formulas 
  and numerical results for the solar mass. In
  the left panel the \texttt{LSP}=$\chi^0$ while in the right panel 
  the \texttt{LSP}=$\tilde{\tau}$. The red (dark) band 
  corresponds to the latest
  neutrino data~\cite{maltoni:2004ei}.}
\label{fig:xx3}
\end{figure}

In the basis where the tree-level neutrino mass matrix is diagonal the
mass matrix at one--loop level can be written as 
\vspace{-1mm}
\begin{equation}
  \widetilde m_\nu= V_\nu^{(0)T} m_\nu V_\nu^{(0)} 
      =\left( \begin{array}{ccc}
      c_1 \widetilde \epsilon_1 \widetilde \epsilon_1 & c_1 \widetilde
      \epsilon_1 \widetilde \epsilon_2  
      & c_1 \widetilde \epsilon_1 \widetilde \epsilon_3 \nonumber\\[+2mm]
      c_1 \widetilde \epsilon_2 \widetilde \epsilon_1 & c_1 \widetilde
      \epsilon_2 \widetilde \epsilon_2  
      & c_1 \widetilde \epsilon_2 \widetilde \epsilon_3 \nonumber\\
      c_1 \widetilde \epsilon_3 \widetilde \epsilon_1 & c_1 \widetilde
      \epsilon_3 \widetilde \epsilon_2  
      & c_0 |\vec\Lambda|^2 
       + c_1 \widetilde \epsilon_3 \widetilde \epsilon_3 
     \end{array} \right) + \cdots 
\label{eq:approx-mat}
\end{equation}
where
\begin{eqnarray}
c_0&=& \frac{M_1 g^2 + M_2 {g'}^2}{4\, \textrm{det}({\cal M}_{\chi^0})} 
 \\[+2mm]
c_1
&=& \frac{3}{16 \pi^2}\, \sin(2\theta_{\tilde b})\, 
m_b\, \Delta B_0^{\tilde b_2\tilde b_1}\ 
\frac{1}{\mu^2}
\end{eqnarray}
The dots in Eq.~(\ref{eq:approx-mat}) correspond to the terms that are
not proportional to the $\tilde \epsilon_i \times \tilde \epsilon_j$
structure, as can be seen from Eq.~(\ref{eq:bot-sbot}). 
Assuming that the bottom-sbottom loop dominates, the $\tilde
\epsilon_i \times \tilde \epsilon_j$ structure is dominant, and the
matrix can be diagonalized approximately under the condition
\begin{equation}
x\equiv\frac{c_1 |\vec{\widetilde{\epsilon}}|^2}{c_0 |\vec \Lambda|^2 }\ll 1
\end{equation}
Then we get
\begin{equation}
  \label{eq:1}
  m_{\nu_2} \simeq \frac{3}{16 \pi^2} \sin(2\theta_{\tilde b}) 
  m_b \Delta B_0^{\tilde b_2\tilde b_1}\ 
  \frac{({\tilde \epsilon}_1^2 + {\tilde \epsilon}_2^2)}{\mu^2}
\end{equation}
and
\begin{equation}
  \label{eq:2}
  \tan^2\theta_\Sol  = \frac{\widetilde  \epsilon_1^2}{\widetilde
    \epsilon_2^2} 
\end{equation}
The results for the masses are presented in Fig.~\ref{fig:xx3}. 
On the left panel we have the data set where the neutralino is the
\texttt{LSP} while on the right panel the \texttt{LSP} is the stau.
We see that the agreement is fairly good, particularly in the region
allowed by the present data.

\subsection{Probing Neutrino Mixing via \texttt{SUSY} Decays}

After having shown, in the previous sections, that the \texttt{BRpV}
model produces an hierarchical mass spectrum for the neutrinos that
can accommodate the present data for neutrino masses and mixings, we
now turn to accelerator physics and will show how the neutrino
properties can be probed by looking at the decays of supersymmetric
particles.

If R-parity is broken the \texttt{LSP} will decay. If the \texttt{LSP}
decays then cosmological and astrophysical constraints on its nature
no longer apply. Thus, within R-parity violating \texttt{SUSY}, a
priori {\em any} superparticle could be the \texttt{LSP}.  In the
constrained version of the \texttt{MSSM} (\texttt{mSUGRA} boundary
conditions) one finds only two candidates for the \texttt{LSP}, namely
the lightest neutralino or one of the right sleptons, in particular
the right scalar tau. Therefore we will consider below these two
cases. However, if we depart from the \texttt{mSUGRA} scenario, it has
been shown recently\cite{hirsch:2003fe} that the \texttt{LSP} can be
of any other type, like a squark, gluino, chargino or even a scalar
neutrino.

\subsubsection{Probing Neutrino Mixing via Neutralino Decays}
\label{sec:neutralino-decays}

\begin{figure}[htbp]
\centering
\begin{tabular}{cc}
\includegraphics[width=0.45\textwidth]{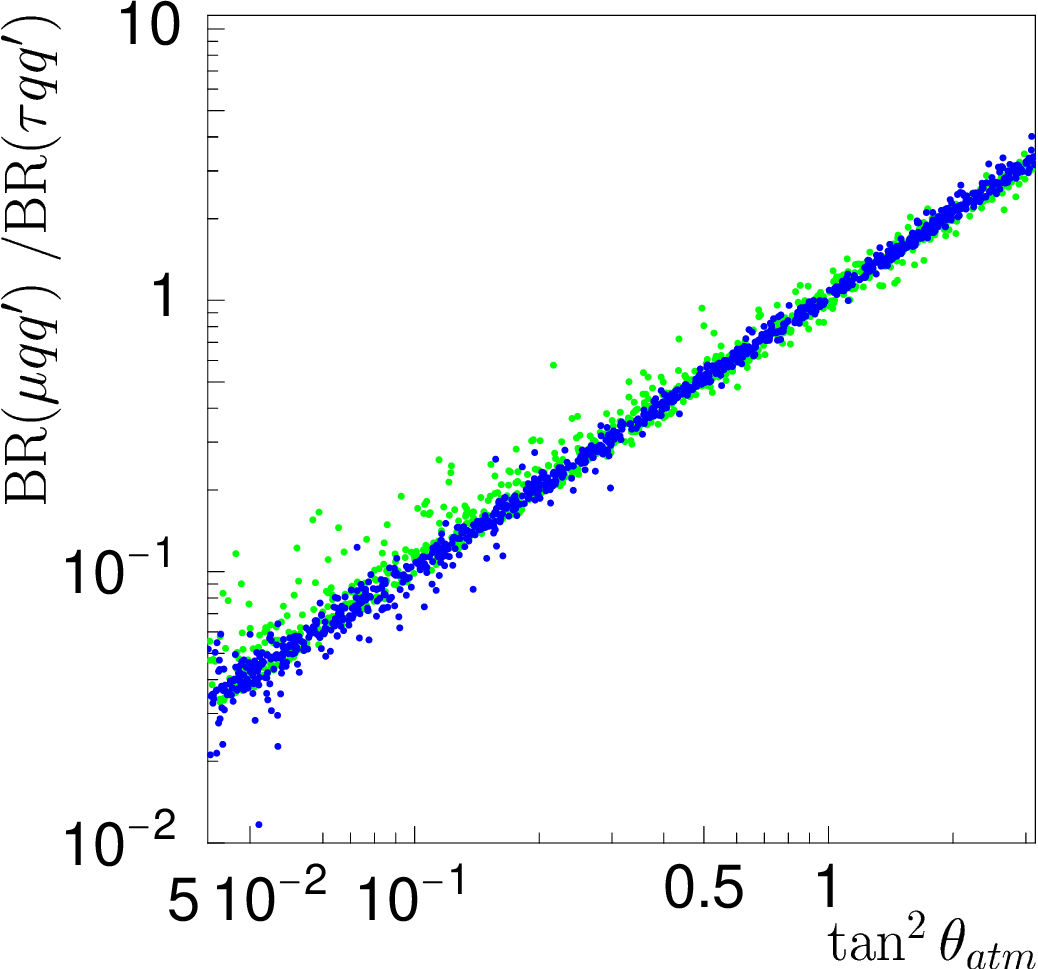}
&
\hskip -1mm
\includegraphics[width=0.45\textwidth]{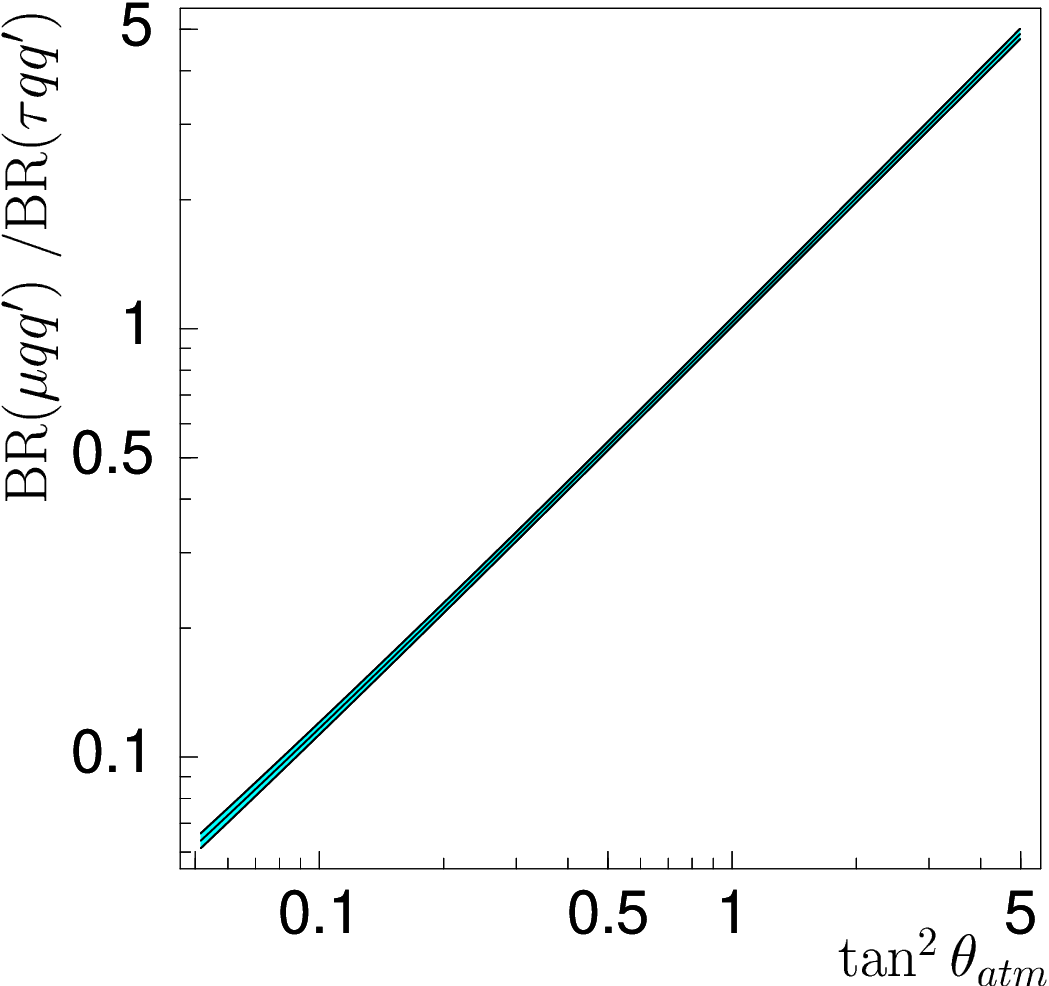}
\end{tabular}
\vspace{-3mm}
\caption{\small
Ratios of semileptonic branching ratios as a function of $\tan^2
\theta_{\Atm}$. On the left for random \texttt{SUSY} values and on the
right for the \texttt{SUSY} point of \Eq{eq:SUSYPOINT}.
}
\label{faro_fig3}
\end{figure}

If the neutralino is the \texttt{LSP} and R-parity is broken, it will
be unstable and it will decay.  It was shown\footnote{The relation of
the neutrino parameters to the decays of the neutralino has also been
considered in Ref.~\cite{Mukhopadhyaya:1998xj}.} in
Ref.~\cite{porod:2000hv}, that the neutralino decays well inside the
detectors and that the visible decay channels are quite large.  We
have shown in the previous sections that the ratios
$|\Lambda_i/\Lambda_j|$ and $|\epsilon_i/\epsilon_j|$ were very
important in the choice of solutions for the neutrino mixing
angles. What is exciting now, is that these ratios can be measured in
accelerator experiments.  In the left panel of Fig.~\ref{faro_fig3} we
show the ratio of branching ratios for semileptonic neutralino decays
into muons and taus: $BR(\chi \to \mu q' \bar q)/ BR(\chi \to \tau q'
\bar q$) as function of $\tan^2 \theta_{\Atm}$. We can see that there
is a strong correlation.  The spread in this figure can in fact be
explained by the fact that we do not know the \texttt{SUSY} parameters
and are scanning over the allowed parameter space. This is illustrated
in the right panel where we considered that \texttt{SUSY} was already
discovered with the following values for the parameters,
\begin{equation}
M_2=120\, {\rm GeV}, \mu=m_0=500\, {\rm GeV}, \tan\beta=5,
 A=-500\, {\rm GeV}.
\label{eq:SUSYPOINT}
\end{equation}

\subsubsection{Probing Neutrino Mixing via Charged Lepton Decays}

\begin{figure}[htbp]
\centering
\begin{tabular}{cc}
\includegraphics[width=0.45\textwidth,height=55mm]{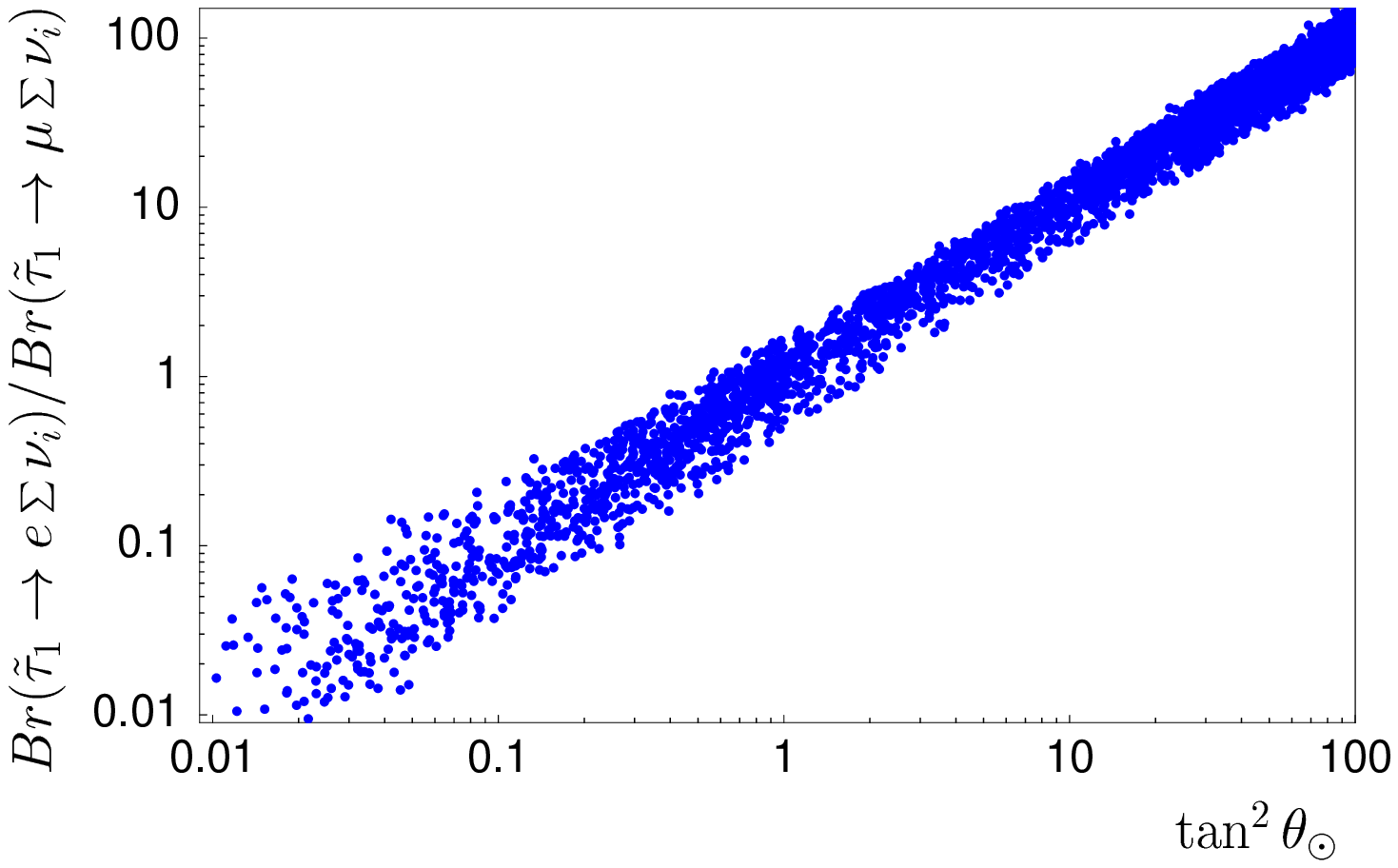}
&
\hskip -3mm
\includegraphics[width=0.45\textwidth,height=55mm]{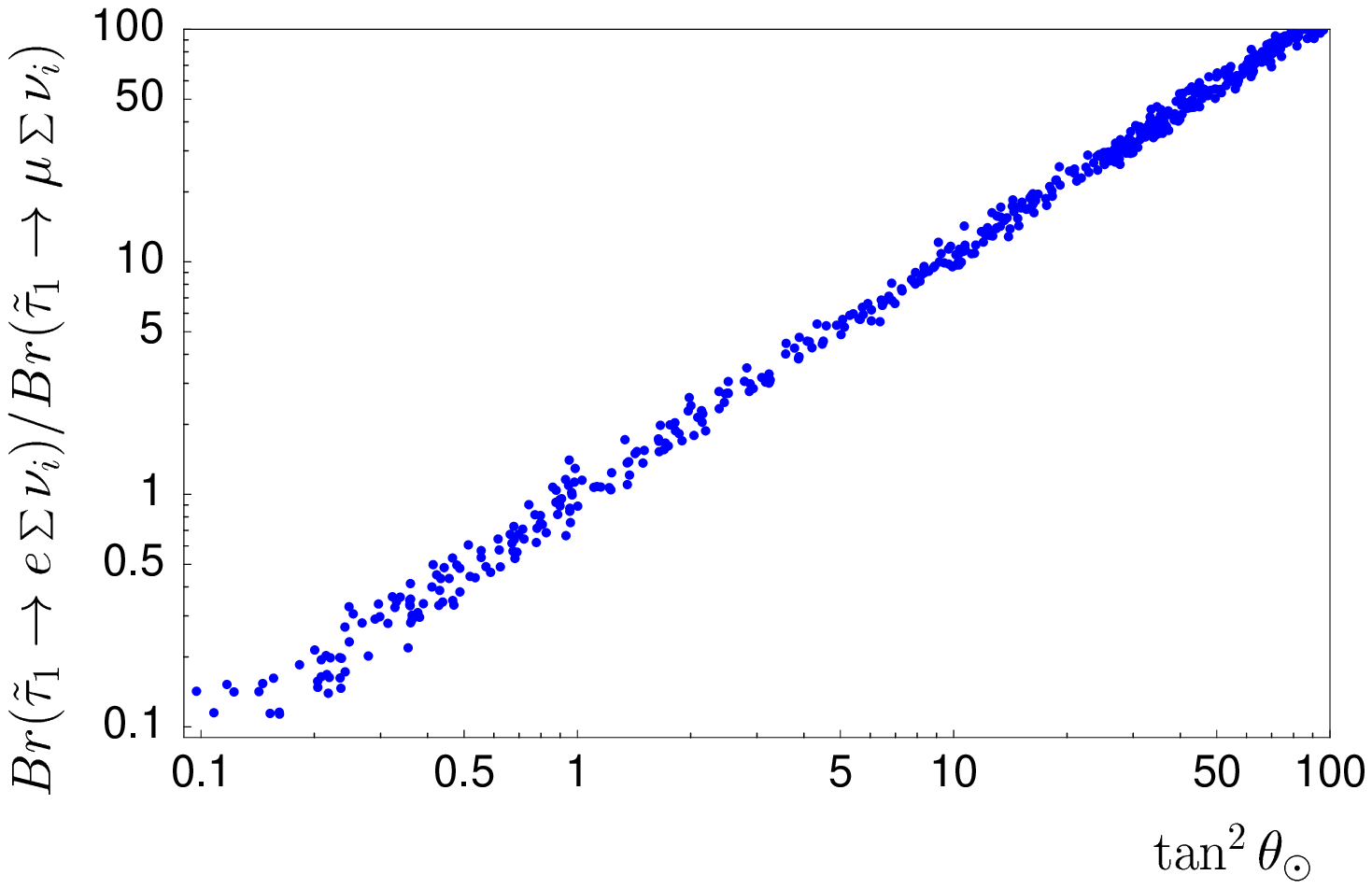}
\end{tabular}
\vspace{-3mm}
\caption{\small
Ratios of branching ratios for scalar tau decays 
  versus $\tan^2\theta_{\odot}$. The left panel shown all data
  points, the right one refers only to data points with
  $\epsilon_2/\epsilon_3$ restricted to the range [0.9,1.1].
}
\label{faro_fig5}
\end{figure}

After considering the case of the \texttt{LSP} being the neutralino,
for completeness, we have also studied~\cite{hirsch:2002ys} the case
where a charged scalar lepton, most probably the scalar tau, is the
\texttt{LSP}.  We have considered the production and decays of $\st$,
$\te$ and $\tm$, and have shown that also for the case of charged
sleptons as \texttt{LSP}s they will decay well inside the detector.
We can correlate branching ratios with neutrino properties.  This is
shown in \Fig{faro_fig5} where the branching ratios of the scalar tau
show a strong correlation with the solar angle.

\subsubsection{Other \texttt{LSP} decays}
\label{sec:otherLSP}

\begin{figure}[htbp]
  \centering
\begin{tabular}{cc}
\includegraphics[width=0.45\textwidth,height=60mm]{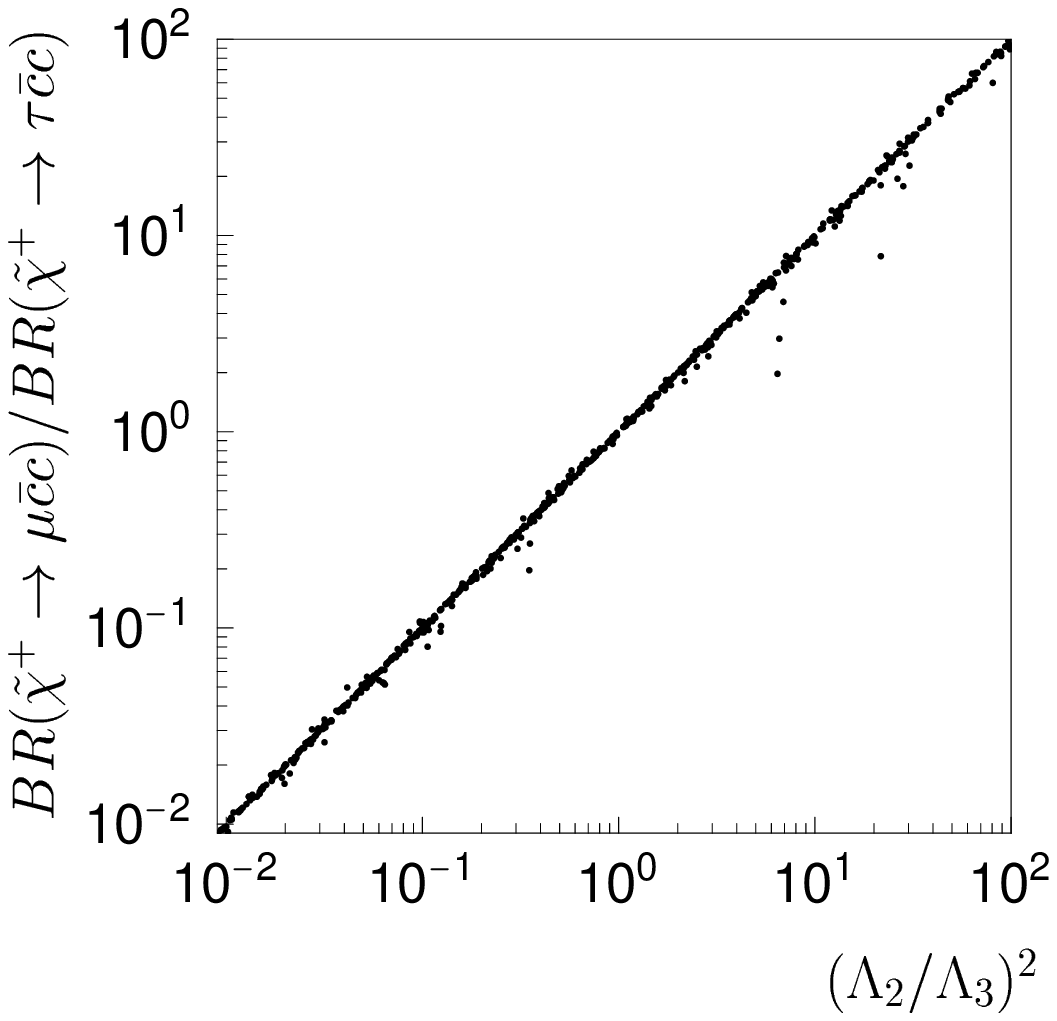}
&
\hskip -3mm
\includegraphics[width=0.45\textwidth,height=60mm]{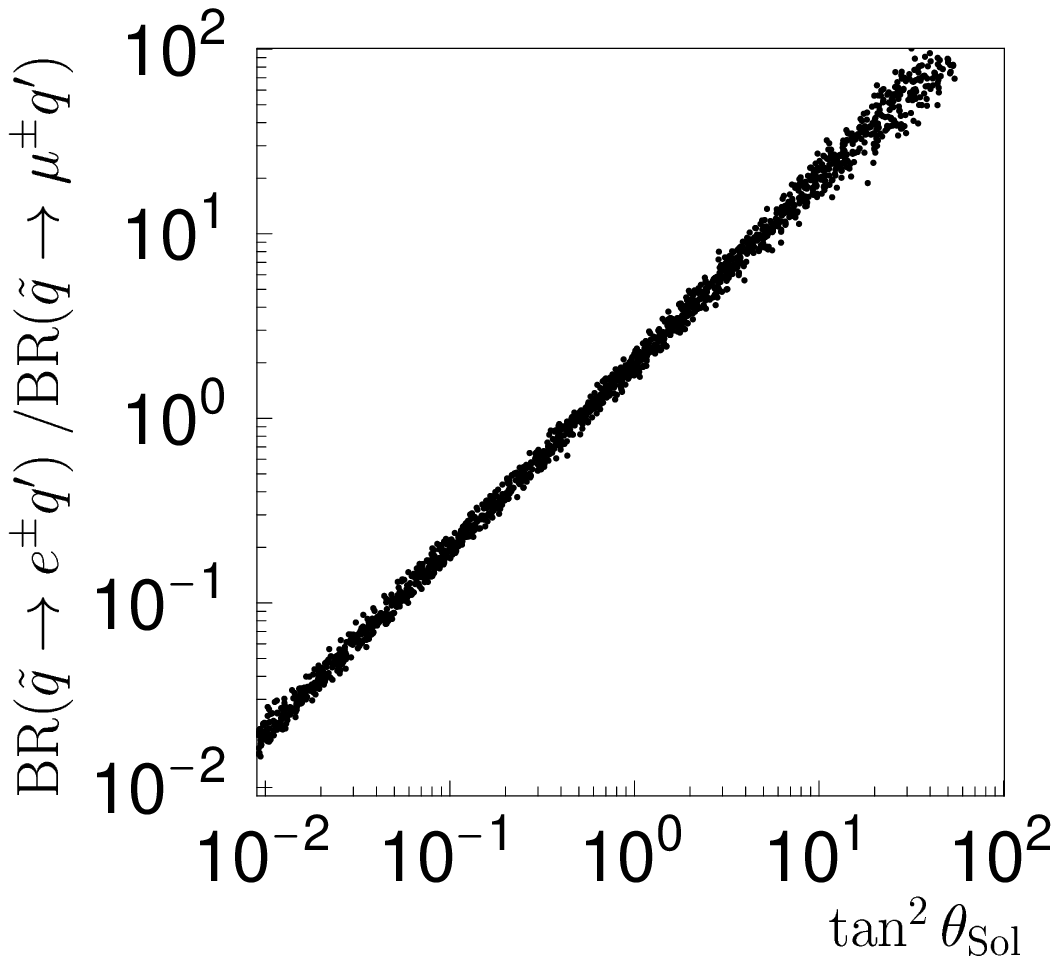}
\end{tabular}
\vspace{-3mm}
\caption{\small
Left panel: ratios of branching ratios for chargino decays 
  versus $(\Lambda_2/\Lambda_3)^2$; Right panel:
ratios of branching ratios for squark decays 
  versus $\tan^2\theta\Sol$. These plots were taken from
  Ref.~\cite{hirsch:2003fe}. }
\label{fig:martin-werner}
\end{figure}

As we discussed in the introduction to this section, if we depart from
the \texttt{mSUGRA} scenario, then the \texttt{LSP} can be of other type,
like a squark, gluino, chargino or even a scalar neutrino, as it has
been shown recently in Ref.~\cite{hirsch:2003fe}. As the decays of
these \texttt{LSP}'s will always depend on the parameters that
violate R-parity and induce neutrino masses and mixings, it is
possible to correlate branching ratios to the neutrino
properties. This is shown in \Fig{fig:martin-werner} taken from
Ref~\cite{hirsch:2003fe}. 
On the left panel we consider the case of chargino decays and show
$BR(\tilde\chi^+ \rightarrow \mu \bar{c} c)/ BR(\tilde\chi^+
\rightarrow \tau \bar{c} c)$ as a function of the ratio
$(\Lambda_2/\Lambda_3)^2$. As we saw, this last ratio is correlated
with the atmospheric angle. So, by looking at the chargino decays one
can test the atmospheric mixing angle. On the right panel we show a
very strong correlation for the solar mixing angle
obtained~\cite{hirsch:2003fe} with the decays of squarks.  Many other
correlations can be obtained making the model over constrained.  So,
in summary, no matter what supersymmetric particle is the
\texttt{LSP}, measurements of branching ratios at future accelerators
will provide a definite test of the \texttt{BRpV} model as a viable
model for explaining the neutrino properties.

\section{Spontaneous Broken R-Parity Model}

\subsection{The Model }

The most general superpotential for the \texttt{SBRp} model is~\cite{masiero:1990uj,romao:1992vu,romao:1992ex,hirsch:2004rw} 
\begin{eqnarray} \nonumber
{\cal W} &\hskip-4mm=\hskip-4mm& \varepsilon_{ab}\Big(
h_U^{ij}\widehat Q_i^a\widehat U_j\widehat H_u^b 
+h_D^{ij}\widehat Q_i^b\widehat D_j\widehat H_d^a 
+h_E^{ij}\widehat L_i^b\widehat E_j\widehat H_d^a \nonumber \\
& &
+h_{\nu}^{ij}\widehat L_i^a\widehat \nu^c_j\widehat H_u^b 
\!- {\hat \mu}\widehat H_d^a\widehat H_u^b 
\!- (h_0 \widehat H_d^a\widehat H_u^b +\delta^2)\widehat\Phi \Big) 
\nonumber \\
& &+\hskip 5mm   h^{ij} \widehat\Phi \widehat\nu^c_i\widehat S_j +
M_{R}^{ij}\widehat \nu^c_i\widehat S_j 
+ \frac{1}{2}M_{\Phi} \widehat\Phi^2 +\frac{\lambda}{3!}
\widehat\Phi^3 
\end{eqnarray}
The first three terms together with the $ {\hat \mu}$ term define the
R-parity conserving \texttt{MSSM}, the terms in the last row only involve the
\21 singlet superfields $({\nu^c}_i,S_i,\Phi)$ carrying a conserved
lepton number assigned as $(-1, 1,0)$, while the remaining terms
couple the singlets to the \texttt{MSSM} fields. To this we will have to add
the soft \texttt{SUSY} breaking terms in the usual
way~\cite{hirsch:2004rw}.

\subsection{Pattern of spontaneous symmetry breaking}

The spontaneous breaking of $R_P$ is driven by nonzero vevs for the
scalar neutrinos. The scale characterizing $R_P$ breaking is set by
the isosinglet vevs,
\begin{equation}
  \label{eq:4}
  \vev{\tilde{\nu^c}} = \frac{v_R}{\sqrt{2}},\quad
  \vev{\tilde{S}}=\frac{v_S}{\sqrt{2}},\quad 
  \vev{\Phi} = \frac{v_{\Phi}}{\sqrt{2}}
\end{equation}
We also have  very small left-handed sneutrino vacuum expectation
values
\begin{equation}
  \label{eq:5}
      \vev{\tilde{\nu}_{Li}} = \frac{v_{Li}}{\sqrt{2}}
\end{equation}
The electroweak breaking is driven by 
\begin{equation}
  \label{eq:6}
 \vev{{H_u}} = \frac{v_{u}}{\sqrt{2}},\quad \vev{{H_d}}=
\frac{v_{d}}{\sqrt{2}}
\end{equation}
with $v^2 = v_u^2 + v_d^2 + \sum_i v_{L i}^2$ and $m_W^2 =
\frac{g^2 v^2}{4}$.
The spontaneous breaking of R--parity also entails the spontaneous
violation of total lepton number. This implies that the majoron
\begin{equation}
  \label{eq:7}
  J=\mathrm{Im} \left[\displaystyle
    \frac{v_L^2}{Vv^2} (v_u H_u - v_d H_d) +
    \sum_i \frac{v_{Li}}{V} \tilde{\nu_{i}} 
    +\frac{v_S}{V} S
    -\frac{v_R}{V} \tilde{\nu^c}\right]
\end{equation}
remains massless, as it is the Nambu-Goldstone boson associated to the
breaking of lepton number.

\subsection{Neutrino-Neutralino-Singlino mass matrix}

For simplicity we choose only one generation of singlet
superfields. Then in the basis $
(-i\lambda',-i\lambda^3,{\tilde H_d},{\tilde H_u},\nu_e,\nu_{\mu},\nu_{\tau},
\nu^c,S,\tilde{\Phi})$ the neutrino/neutralino mass matrix is $10\times10$,
\begin{equation}
  \label{eq:8}
  \mathbf{M_N}=
  \left(\begin{array}{lllll}
      \mathbf{M_{\chi^0}} & \mathbf{m_{\chi^0\nu}}& \mathbf{m_{\chi^0\nu^c}}& 
      \mathbf{0}& \mathbf{m_{\chi^0\Phi} } \\
      \\
      \mathbf{m^T_{\chi^0\nu}} & \mathbf{0} & \mathbf{m_{D}} & 
      \mathbf{0} & \mathbf{0} \\
      \\
      \mathbf{m^T_{\chi^0\nu^c}}&\mathbf{m^T_{D}} & \mathbf{0} &
      \mathbf{M_{\nu^c S}} & \mathbf{M_{\nu^c\Phi}} \\
      \\
      \mathbf{0} &\mathbf{0} &
      \mathbf{M^T_{\nu^c S}} &\mathbf{0} &\mathbf{M_{S\Phi}} \\
      \\
      \mathbf{m^T_{\chi^0\Phi} } & \mathbf{0} & \mathbf{M^T_{\nu^c\Phi}} & 
      \mathbf{M^T_{S\Phi}} & \mathbf{M_{\Phi}}
    \end{array} \right)
\end{equation}
where 
$\mathbf{M_{\chi^0}}$ is the usual \texttt{MSSM} neutralino mass
matrix, and
\begin{equation}
  \label{eq:9}
  \mathbf{m^T_{\chi^0\nu}} \hskip-1mm=\hskip-1mm
  \left(\! \begin{array}{llll}
      -\frac{1}{2}g'v_{L e} &\frac{1}{2}gv_{L e} & 0 & \epsilon_e \\[2mm]
      -\frac{1}{2}g'v_{L \mu}& \frac{1}{2}gv_{L \mu}& 0& \epsilon_{\mu}\\[2mm]
      -\frac{1}{2}g'v_{L \tau} & \frac{1}{2}gv_{L \tau} & 0& \epsilon_{\tau} 
    \end{array}\! \right)
 \end{equation}
\begin{equation}
  \label{eq:10}
  (\mathbf{m_{D}})_{i} = \frac{1}{\sqrt{2}}h_{\nu}^{i}v_u
 \end{equation}
with an effective ``bilinear R--parity'' parameter
\begin{equation}
  \label{eq:11}
  \epsilon_i = \frac{1}{\sqrt{2}}h_{\nu}^{i} v_R
\end{equation}
The other block matrices in Eq.~(\ref{eq:8}) can found explicitly in
Ref.\cite{hirsch:2004rw}.

%%%-------------------- SLIDE -----------------------
\subsection{The effective neutrino mass matrix}

The exact diagonalization of Eq.~(\ref{eq:8}) can only be done
numerically. However due to the smallness of the neutrino masses, an
effective neutrino mass 
matrix can be cast into a very simple form 
\begin{equation}
  \label{eq:12}
  (\mathbf{m_{\nu\nu}^{\rm eff}})_{ij} = a \Lambda_i \Lambda_j + 
  b (\epsilon_i \Lambda_j + \epsilon_j \Lambda_i) +
  c \epsilon_i \epsilon_j.      
\end{equation}
where the model dependence in hidden in the parameters $a,b,c$.  The
effective bilinear R--parity violating parameters $\epsilon_i$ and
$\Lambda_i$, are given in Eq.~(\ref{eq:11}) and Eq.~(\ref{eq:lambda}),
respectively.  This equation resembles very closely the result for the
\texttt{BRpV} model once the dominant 1-loop corrections are taken
into account.
The tree-level result of the explicit bilinear model can be recovered
in the limit $M_R,M_{\Phi} \to \infty$.
\begin{equation}
  \label{eq:13}
  a = \displaystyle \frac{g^2M_1
    +g'^2 M_2}{4{\rm Det}\mathbf{M_{\chi^0}}},\quad b=0, \quad c=0
\end{equation}
In this limit only one non-zero neutrino mass remains.  The relative
size of the coefficient $c$ compared to the corresponding 1-loop
coefficient dictates if the 1-loop corrections or the contribution
from the singlet fields are more important.  Both extremes can be
realized in our model. Large branching ratios of the Higgs into
invisible final states require sizable values of $h$ and $h_0$.  Then
the ``singlino'' contribution dominates.

\subsection{The Neutral Scalar Bosons Mass Matrix}

The $8\times 8$ scalar mass matrix is a symmetric
matrix that in the basis
$(H_d^0,H_u^0,\tilde\nu_i,\Phi,\tilde{S},\tilde\nu^c)$
 takes the form,
 \begin{equation}
   \label{eq:14}
  M^{S^2}=\left[
    \begin{array}{lll}
      M^{S^2}_{HH}
      & 
      M^{S^2}_{H\widetilde L} 
      & 
      M^{S^2}_{HS}\\[+4.5mm]
      M^{S^2}_{H\widetilde L}\!{}^T 
      & 
      M^{S^2}_{\widetilde L \widetilde L} 
      & 
      M^{S^2}_{\widetilde L S}\\[+4.5mm]
      M^{S^2}_{HS}\!{}^T 
      &
      M^{S^2}_{\widetilde L S}\!{}^T 
      & 
      M^{S^2}_{SS}
    \end{array}
    \right]
 \end{equation}
written in terms of block matrices that can be found in
Ref.\cite{hirsch:2004rw}. The pseudo--scalar mass matrix has a similar
form. In practice they are quite complicated and have to be diagonalized
numerically. The most important feature of the pseudo--scalar mass
matrix is the existence of two massless Goldstone bosons,
corresponding to the spontaneous breaking of gauge symmetry (the usual 
Goldstone $G_0$ ``eaten'' by the $Z^0$) and to the spontaneous breaking of
R--parity, the majoron $J$.
In the basis 
\begin{equation}
  \label{eq:16}
A'_0=(H_d^{0 I},H_u^{0 I},\tilde{\nu}^{1 I},\tilde{\nu}^{2 I},
\tilde{\nu}^{3 I},\Phi^I,\tilde{S}^I,\tilde{\nu}^{c I})  
\end{equation}
 we have
\begin{eqnarray}
  \label{eq:15}
  G_0&=&(N_0\, v_d,-N_0\,
    v_u,N_0\,v_{L1},N_0\, v_{L2},N_0\, v_{L3},0,0,0)\nonumber \\[+2mm]
J&=&(-N_1 v_d,N_1 v_u, N_2 v_{L1}, N_2
  v_{L2}, N_2 v_{L3},0,N_3 v_S,-N_3 v_R)
\end{eqnarray}
where the $N_i$ are normalization constants.
It can easily be checked that they are orthogonal $G_0 \cdot J=0$.

\subsection{Higgs Boson Production}

Supersymmetric Higgs bosons can be produced at an $e^+
e^-$ collider via the so--called Bjorken process,
\begin{equation}
  \label{HZZ1}
  {\cal L}_{HZZ}
  =\displaystyle
  \sum_{i=1}^8 (\sqrt 2 G_F)^{1/2} M_Z^2 Z_{\mu}Z^{\mu} \eta_{i} H_i
\end{equation}
where the $\eta_i$ are combinations of the doublet scalars vevs and
rotation matrices:
\begin{equation}
  \label{eq:18}
  \eta_i= \frac{v_d}{v} R^S_{i 1}
  + \frac{v_u}{v} R^S_{i 2} 
  + \sum_{j=1}^3 \frac{v_{Lj}}{v} R^S_{i j+2}
\end{equation}
In comparison with the \texttt{SM} the coupling of the lightest CP--even Higgs
boson to the $Z$ is reduced by a factor
\begin{equation}
  \label{eq:19}
  \eta \equiv \eta_1 \leq 1 .
\end{equation}
In the \texttt{MSSM} we have $\eta=\sin(\beta-\alpha)$.

\subsection{Higgs Boson Decay}

We are interested here in the ratio

\begin{equation}
  \label{eq:ratio}
  R_{Jb}=\displaystyle
  \frac{\Gamma(h \to JJ)}{\Gamma(h \to b \bar{b})}
\end{equation}
of the invisible decay to the SM decay into b-jets. 
These decay widths are easily obtained~\cite{hirsch:2004rw},

\begin{equation}
  \label{eq:JJ}
  \Gamma(h\to  JJ)=\displaystyle
  \frac{g_{hJJ}^2}{32\pi m_h}
\end{equation}
and
\begin{equation}
    \label{eq:bb}
    \Gamma(h\to  b \bar{b})=\displaystyle
    \frac{3 G_F \sqrt{2}}{8\pi}\,
    \left(R^S_{11}
        \right)^2\, m_h\,  m_b^2 \left[ 1-4
  \left(\frac{m_b}{m_h}\right)^2 \right]^{3/2}
\end{equation}
We want to look at situations where $R_{Jb} > 1$, but that at the same
time the Higgs boson is produced at a reasonable rate, which means that
$\eta$ parameter should not be too small.

\subsection{Scanning Strategy}

As the model has a large number of parameters we have to follow a
scanning strategy that can put in evidence the possibility of the Higgs
boson decaying invisibly. This is achieved by fixing as many
parameters as we can. Without loss of generality we then fix the
\texttt{MSSM} parameters to the \texttt{SPS1a} benchmark point:
\begin{eqnarray}
  \label{eq:23}
&&      m_0 = 100 \rm{GeV}, \quad m_{1/2} = 250 \rm{GeV}, \quad
\tan\beta=10 \nonumber \\[+2mm] 
&& A_0 = -100 \rm{GeV},  \quad \mu <0
\end{eqnarray}
For the singlet parameters we choose to start with
\begin{eqnarray}
  \label{eq:24}
  &&   v_R= v_S= v_{\Phi} = -150 \mathrm{GeV} \nonumber \\[+2mm] 
  && M_R = - M_{\Phi} = \delta = 10^3 \mathrm{GeV} \nonumber \\[+2mm] 
  &&h=0.8\ , h_0 = -0.15\ , \lambda=0.1
\end{eqnarray}
and then perform variations around these values in a controlled way.
Finally the explicit bilinear parameters, $\epsilon_i$ and
$\Lambda_i$, are fixed approximately such that neutrino masses
and mixing angles are in agreement with experimental data.

%%%-------------------- SLIDE -----------------------

\subsection{Numerical results: General Superpotential}

\begin{figure}[htbp]
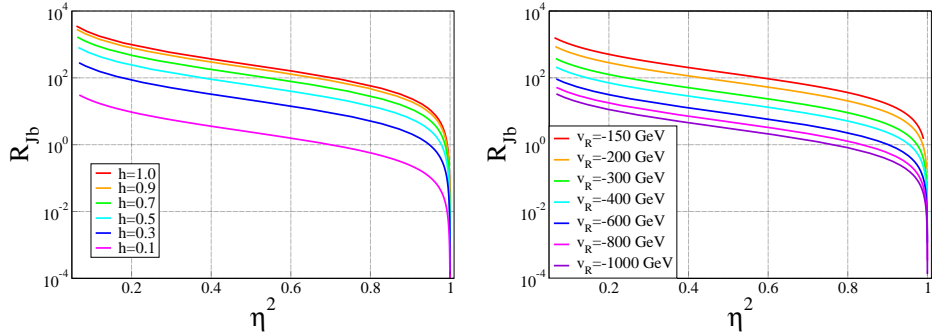

  \centering
  \begin{tabular}{cc}
  \includegraphics[clip,width=0.47\linewidth]{plot-54-i.eps}
  &
  \includegraphics[clip,width=0.47\linewidth]{plot-59-i.eps}      
  \end{tabular}
  \caption{$R_{Jb}$ as a function of $\eta^2$: a) For constant values
  of $h$. b) For constant values of $v_R=v_S$.}
  \label{fig:xxx}
\end{figure}

In Fig.~\ref{fig:xxx} we show the plot of $R_{Jb}$ as a function of
$\eta^2$. On  the left panel we kept $h$ constant and varied
continuously the other singlet parameters, while on the right panel
the vevs were kept. We see that it is indeed possible to achieve,
\textit{at the same time}, both large branching ratio into the
invisible mode ($h\to JJ$), and large production cross sections.
In Ref.\cite{hirsch:2004rw} a complete analysis of the parameter space
where this happens is presented.

%%%-------------------- SLIDE -----------------------

\section{Conclusions}

The \texttt{BRpV} model is a simple extension of the \texttt{MSSM}
that leads to a very rich phenomenology.  We have calculated the
one--loop corrected masses and mixings for the neutrinos in a
completely consistent way, including the RG equations and correctly
minimizing the potential.  We have shown that it is possible to get
easily maximal mixing for the atmospheric neutrinos and large angle
MSW, as it is preferred by the present neutrino data. We have also
obtained approximate formulas for the solar mass and solar mixing
angle, that we found to be very good, precisely in the region of
parameters favored by this data.

We emphasize that the \texttt{LSP} decays inside the detectors, thus
leading to a very different phenomenology than the \texttt{MSSM}.  No
matter what supersymmetric particle is the \texttt{LSP} measurements
of branching ratios at future accelerators will provide a definite
test of the \texttt{BRpV} model as a viable model for explaining the
neutrino properties.

We have also discussed the possibility of invisibly decaying Higgs
boson in the context of the \texttt{SBRp} model.  One of the neutral
CP-odd scalars in this model, is the Nambu-Goldstone boson associated
to the breaking of lepton number, the majoron $J$.  Neutrino masses
and mixings can easily be explained.  In contrast to the
\texttt{MSSM}, where the Higgs boson can decay invisibly only to
supersymmetric states, in our case the Higgs boson can decay mainly
invisibly through the decay $h\to JJ$.  As a result, our analysis
indicates that invisibly decaying Higgs bosons should be an important
topic in the agenda of future accelerators, such as the Large Hadron
Collider and the Next Linear Collider.

%\bibliographystyle{h-physrev4} 
%\bibliography{romao-ref}

\end{document}